\documentclass[aps,prl,reprint,showpacs,preprintnumbers,amsmath,amssymb,graphicx,
superscriptaddress,10pt]{revtex4-1}
\usepackage{epsfig}
\usepackage{dcolumn}
\usepackage{bm}
\usepackage{natbib}
\usepackage{graphicx}
\usepackage{subfigure}
\usepackage{sidecap}

\begin{document}

\author{G.\ Karapetrov}\email{e-mail: goran@drexel.edu}
\affiliation{Department of Physics, Drexel University, 3141
Chestnut Street, Philadelphia, PA 19104, USA}

\author{V.\ Yefremenko}
\affiliation{Materials Science Division, Argonne National
Laboratory, Argonne, Illinois 60439, USA}

\author{G.\ Mihajlovi\'c}\altaffiliation[Present
address: ]{San Jose Research Center, Hitachi Global Storage
Technologies, San Jose, CA 95135} \affiliation{Materials Science
Division, Argonne National Laboratory, Argonne, Illinois 60439,
USA}

\author{J.\ E.\ Pearson}
\affiliation{Materials Science Division, Argonne National
Laboratory, Argonne, Illinois 60439, USA}

\author{M.\ Iavarone}
\affiliation{Department of Physics, Temple University,
Philadelphia, PA 19122, USA}

\author{V.\ Novosad}
\affiliation{Materials Science Division, Argonne National
Laboratory, Argonne, Illinois 60439, USA}

\author{S.\ D.\ Bader}
\affiliation{Materials Science Division, Argonne National
Laboratory, Argonne, Illinois 60439, USA}

\title{Evidence of Vortex Jamming in Abrikosov Vortex Flux Flow Regime}
\date{\today}
\tighten

\begin{abstract}
We report on dynamics of non-local Abrikosov vortex flow in
mesoscopic superconducting Nb channels. Magnetic field dependence
of the non-local voltage induced by the flux flow shows that
vortices form ordered vortex chains. Voltage asymmetry
(rectification) with respect to the direction of vortex flow is
evidence that vortex jamming strongly moderates vortex dynamics in
mesoscopic geometries. The findings can be applied to
superconducting devices exploiting vortex dynamics and vortex
manipulation, including superconducting wires with engineered
pinning centers.
\end{abstract}

\date{\today}
 \maketitle

The dynamic behavior of vortices in type-II superconductors is
important in applications~\cite{campbell_evetts} and provides an
exemplary model system. Abrikosov vortices in type-II
superconductors are set in motion when the Lorentz force due to
local supercurrents exceeds the strength of vortex pinning forces,
resulting in energy dissipation. Therefore, control of the dynamic
behavior of vortices has a broad relevance for reducing losses in
superconducting wires and microwave devices. Also, the ability to
control and manipulate single magnetic flux quanta could lead to
novel superconducting devices, such as current
rectifiers~\cite{VVM_ratchet,plourde_ratchet1, plourde_ratchet2},
logic elements~\cite{hastings,puig_automata}, or single photon
detectors~\cite{kadin_prl}.

In this work we demonstrate vortex jamming effect through
long-range manipulation of Abrikosov vortices in mesoscopic
superconducting wires. Non-local vortex flow is initiated by
applying a Lorentz force on parts of the vortex lattice, while
observing vortex motion in areas where no external force is
present. The nonlocal flow is due to the finite elasticity of the
vortex lattice, {\em i.e.} finite tilt modulus
$c_{44}$~\cite{giaver}, compression $c_{11}$ and shear $c_{66}$
moduli~\cite{grigorieva,kokubo,kokubo_channel}. Enhanced rigidity
of the vortex lattice ({\em i.e.} enhanced $c_{11}$ and $c_{66}$)
in mesoscopic superconductors with reduced dimensionality
(superconducting strips) leads to the possibility of non-local
manipulation of vortices on the order of several hundred vortex
lattice spacings~\cite{grigorieva,vodolazov,helzel}. Non-local
vortex manipulation opens opportunities to sense perturbations of
the vortex system far from the point where the external force acts
on an individual vortex. This dramatically increases the
sensitivity to dynamic properties of Abrikosov vortices in
mesoscopic superconductors.

Here we show that long-range manipulation of vortices in
low-$\kappa$ materials, such as Nb, can be exploited to study
granular-like properties of vortex matter, including vortex
jamming~\cite{reichhardt_jamming_prb,reichhardt_jamming_physicac}.
Reduction or complete termination of vortex flow due to strong
vortex-vortex interaction in constricted (funnel) geometries leads
to a reduction of dissipation due to vortex movement. Similar to
granular materials, the affinity to jam in funnel geometries is
most pronounced at high particle densities, {\em i.e.} in our case
at high magnetic fields. We examine the jamming effect in the case
of a single constriction that is much larger than the size of a
vortex. Vortices are channelled into a constriction on the order
of few vortex cores wide~\cite{reichhardt_jamming_physicac}. We
show that the vortex mobility is asymmetric with respect to the
direction of motion of the vortices, as observed by the nonlocal
voltage in Nb bridges, and discussed theoretically by Vodolazov et
al.~\cite{vodolazov}. The experimental evidence of granular-like
behavior of vortex matter that we show in this paper should
initiate wider application of theoretical concepts developed in
the physics of granular materials to the behavior of
superconducting vortices. Based on these concepts, the engineering
of pinning landscapes that suppress vortex motion via jamming will
lead to lower dissipation at high magnetic fields in a wide range
superconducting devices and wires.

\begin{figure}
\includegraphics[width=3.0in]{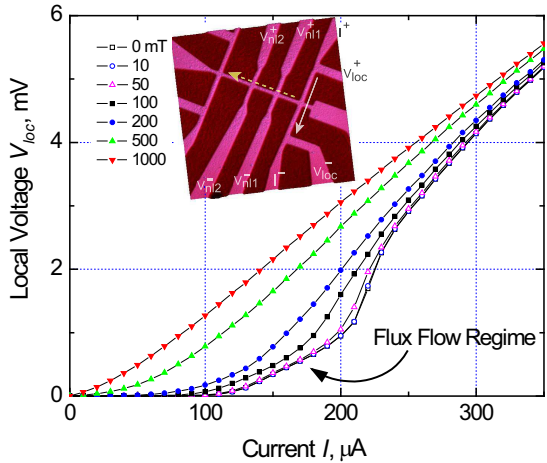}
\caption{(Color online) Local current-voltage characteristics of
the Nb bridge at 2 K. Inset: atomic force microscopy image of the
device used in the studies of nonlocal vortex movement - {\em I} -
local current, $V_{loc}$ - local voltage, $V_{nl1}$ and $V_{nl2}$
- nonlocal voltage leads. The distance between {\em I} and
$V_{nl1}$ and $V_{nl2}$ at the central vortex guide channel is 1.5
$\mu$m and 3 $\mu$m, respectively.} \label{Figure1}
\end{figure}

The 100-nm-thick Nb film was sputtered using a $dc$ magnetron on a
silicon wafer covered with 500 nm of silicon nitride. The critical
temperature of the as-grown, high quality Nb film was 9.1 K with
transition width of 50mK and residual resistivity ratio RRR$sim$5.
We used e-beam lithography and reactive ion etching (combination
of CF$_4$ and SF$_6$ plasma) to define structures similar to those
shown in the inset of Figure~\ref{Figure1}. The 200-nm wide bridge
had the normal state resistance of 8.65 $\mu\Omega$cm. Considering
that the slope dH$_{c2}$/dT near T$_c$ is 0.4 T/K and the fact
that we are in the dirty limit,  we obtain the basic
Ginzburg-Landau superconducting parameters of the structure: the
coherence length $\xi$(0) = 13.5 nm and the penetration depth
$\lambda$(0) = 145 nm, resulting in $\kappa$=
$\lambda$(0)/1.63$\xi$(0)=6.6~\cite{kes_tsuei}.

The choice of material was motivated by the requirement of
strongly interacting vortex matter in order to reduce the
compressibility of the lattice and enhance the vortex jamming
effects. This can be accomplished by enhancement of the elasticity
moduli of the vortex lattice, which led us to use type-II
superconductors with small values of $\kappa=\lambda/\xi$
~\cite{campbell_jpc1,*campbell_jpc2,*brandt_rpp,*blatter_bible}.
However, the intrinsic pinning in such thin films is strong due to
the same reasons, and the vortex-pin interaction usually dominates
over the vortex-vortex interaction. Therefore, the quality of the
Nb thin film as well as precise definition of the mesoscopic
structures are both important.

The Nb films were patterned into multi-terminal structures, as
shown in the inset of Fig.~\ref{Figure1}. Direct current was
applied through terminals {\em I}, and a local voltage was
measured via terminals $V_{loc}$ (conventional four-terminal
geometry), while the non-local voltage was measured via terminals
$V_{nl1}$ and $V_{nl2}$ using a Keithley 2182 nanovoltmeter. The
distance between the local leads to the $V_{nl1}$ is 1.5 $\mu$m
and to the $V_{nl2}$ is 3.0 $\mu$m. The center superconducting
line with width {\em w} = 200 nm serves as a guiding channel for
Abrikosov vortices. When the external magnetic field is applied
perpendicular to the plane of the film, vortices are nucleated in
the superconducting pads located at either side of the sample, and
depending on the direction of the current and the polarity of the
vortices, the vortices can be moved either to the left or to the
right side of the sample under the influence of Lorentz force. The
coupling of the left and right superconducting Nb pads with the
vortex guiding channel is through asymmetric constrictions. In
case of granular-like behavior of the vortices, the funnel
provides a well-defined guided concentrator for vortex movement to
the right, facilitating vortex
jamming~\cite{reichhardt_jamming_prb}.

The geometry of the structure is such that the applied current is
confined to the current lead {\em I} and exerts a direct force on
the vortices that are only in that part of the structure. The
applied current decays exponentially $\propto\exp(-\pi x/w)$ in
the vortex guiding channel at a distance {\em x} away from the
intersection between the local current lead and the vortex
channel~\cite{grigorieva}. Therefore, no voltage is expected at
both $V_{nl1}$ and $V_{nl2}$ when the structure is either in the
normal or in superconducting state without vortex flow.

Local current-voltage characteristics at 2 K in an applied
magnetic field show that we can isolate the vortex flux-flow
regime in the range of currents between 100-200 $\mu$A
(Fig.~\ref{Figure1}). The flux flow regime in this part of the
vortex phase diagram is characterized by high vortex flow
velocities~\cite{peroz}. The onset of the linear regime at
$\sim$~100 $\mu$A can be interpreted as dynamic phase
transition~\cite{koshelev_dynmelt} from plastic flow at I
$\lesssim$ 100 $\mu$A to a coherent movement of the vortex lattice
as a whole at I $\gtrsim$ 100 $\mu$A.

\begin{figure}
\includegraphics[width=3.0in]{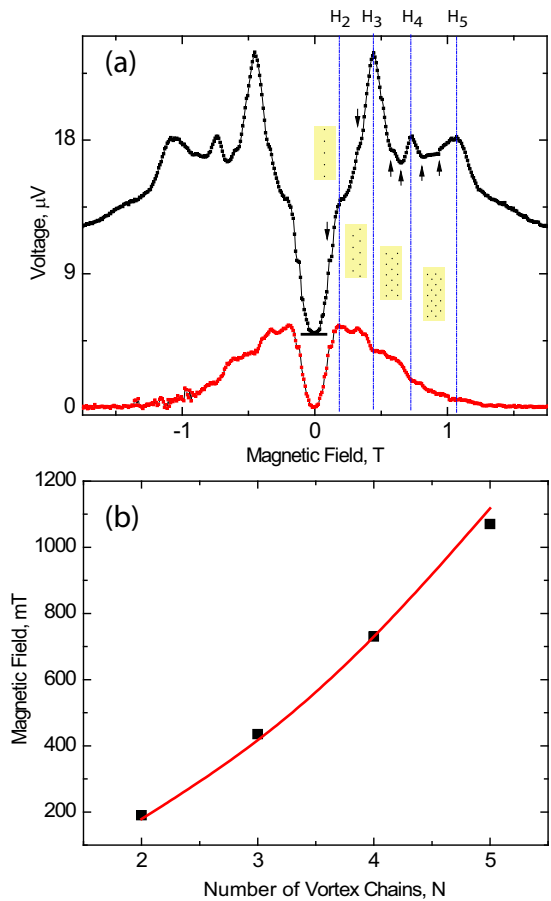}
\caption{(Color online) (a) Nonlocal voltage as a function of
applied magnetic field measured in the 200-nm-wide channel at a
distance of 1.5 $\mu$m ($V_{nl1}$ - black) and 3 $\mu$m ($V_{nl2}$
- red) from the applied current. The curves are shifted vertically
for clarity. Blue vertical dotted lines correspond to magnetic
fields at which vortex chain transitions take place with schematic
of vortex chain configurations at different field intervals.
Arrows denote additional kinks in $V_{nl}(H)$ dependence; (b)
magnetic field dependence of the vortex chain transitions in the
channel showing experimental values (points) and theoretical curve
from Eq. (1). No fitting parameters are used.}\label{Figure2}
\end{figure}

The magnetic field dependence of the nonlocal voltage is shown in
Fig.~\ref{Figure2}a. The current of 150 $\mu$A was applied through
the structure in order to initiate the flux flow through the
superconducting vortex guiding channel. The nonlocal voltage
appears at nonzero magnetic field, reaches some finite value and
then disappears at higher magnetic fields approaching H$_{c2}$ at
this temperature. The voltage signal shown in this figure is
obtained by taking a difference in voltage signals for opposite
directions of the applied current thus eliminating any thermal
emf. The signal is symmetric about zero magnetic field. With
increasing distance from the driving current the non-local voltage
is reduced, but it is still finite even 3 $\mu$m away. The
non-local voltage decays with distance from the current lead where
Lorentz force is present. If the vortex lattice was ideally rigid
the non-local voltage would not decay with distance. But finite
rigidity of the vortex lattice causes local compression of the
lattice and presence of disorder in the film causes finite
pinning. Both of these contribute to loss of vortex flow along the
vortex carrying stripe and decay of non-local voltage with
distance.

One stark difference from earlier reports on the nonlocal vortex
motion is the appearance of highly symmetric peaks and kinks in
the nonlocal voltage at specific values of the magnetic field. The
features are correlated between the first and second nonlocal
probe, indicating highly correlated, long-range vortex flow along
the channel. The main features in the nonlocal voltage curves can
be observed at 190, 435, 730, and 1070 mT. Weakly pronounced
features, such as kinks, are present at intermediate fields, and
they correlate between the two nonlocal voltage signals.
Considering the mesoscopic size of the vortex channel the main
peaks are to be associated with transitions between structural
configuration of the moving vortex lattice in the channel. The
narrow width of the channel that is on the order of the
vortex-vortex spacing imposes its 1D geometry on the vortex
lattice. A triangular vortex lattice re-organizes into a vortex
chain
structure~\cite{ivlev_tsf,*ivlev,guimpel,takacs,*carter,*carneiro,*luzhbin,*field_gelfand,*karapetrov_PRL,*karapetrov_PRB2009}.
Since the driving current is strong, such that the vortex lattice
is moving with relatively high speed, the vortex lattice can be
considered dynamically ordered~\cite{koshelev_dynmelt} and its
configuration can be calculated by applying equilibrium
conditions. At fields below H$_2$ the vortices are arranged in a
regular chain moving with steady velocity along the center axis of
the mesoscopic channel. A Bean-Livingston barrier aligns the
vortices in the center of the channel. When the magnetic field is
increased, the intervortex spacing in the chain reduces gradually,
until the intervortex distance becomes on the order of the half of
 strip width. At field H$_2$ a structural transformation occurs
- the vortex chain is suddenly split into two chains and the
intervortex distance within each chain increases. This leads to a
rapid increase of the total amount of vortices flowing through the
channel leading to a peak in nonlocal voltage. Further increase in
magnetic field preserves the two-chain configuration while
gradually decreasing the intervortex spacing within each of the
chains. The process of chain splitting repeats at H$_3$ and H$_4$
with formation of three and four vortex chains, respectively with
a corresponding peak in the nonlocal voltage. Ginzburg- Landau
calculations~\cite{ivlev} show that the magnetic fields at which
the equilibrium reordering of the vortex lattice occurs are:
\begin{eqnarray}
&& \mu_0H_N=\frac{\sqrt{3}}{2}\frac{\Phi_0
\lambda_{ab}}{\lambda_c}\left( \frac{N}{w}\right)^2 \label{gl1}
\end{eqnarray}
where $\lambda_{ab}$ and $\lambda_{c}$ are the in-plane and
out-of-plane penetration depths, respectively, $\Phi_0$ is the
magnetic flux quantum, {\em w} is the width of the strip, and
N=2,3,4... is the peak index. Since Nb film is in dirty
superconducting limit, the anisotropy
$\lambda_{ab}/\lambda_{c}~1$. The correspondence of the peaks to
the theoretical model is shown in Fig.~\ref{Figure2}b. The slight
discrepancy at higher order peaks might be due to the
renormalization of the Bean-Livingston barrier at high fields,
which would result in a slight change of the effective vortex
channel width {\em w}.

\begin{figure}
\includegraphics[width=3.5in]{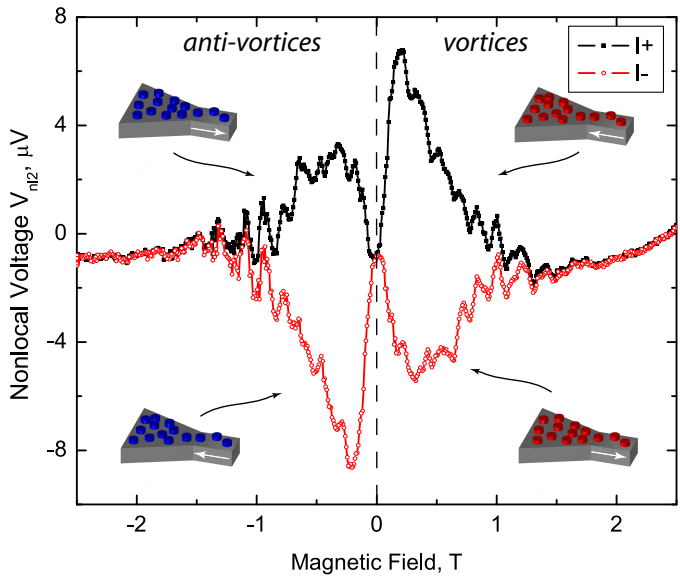}
\caption{(Color online) Non-local voltage $V_{nl2}(I,H)$ as a
function of the applied magnetic field for two opposite directions
of the driving current (T=2.4~K). Vortices of opposite polarities
exhibit the same rectification behavior when moving in the same
direction. } \label{Figure3}
\end{figure}

Additional more weakly pronounced kinks in the nonlocal voltage
can be observed at magnetic fields that are between the H$_N$
vortex transitions. In order to elucidate the nature of these
additional features we go back to the origin of the nonlocal
voltage. The nonlocal voltages V$_{nl1}$ and V$_{nl2}$ shown in
Figure~\ref{Figure2}a are the average absolute voltages induced by
positive and negative driving currents $I^+$ and $I^-$ in the
current channel, i.e.
$V_{nl}(H)$=$\frac{V_{nl}(I^+,H)-V_{nl}(I^-,H)}{2}$. The $V_{nl1}$
and $V_{nl2}$ signals are dominated by the vortex flux flow in the
central channel and any asymmetries of vortex motion up and down
the channel are averaged out. In Fig.~\ref{Figure3} we show the
traces of the original $V_{nl2}(I^+,H)$ and $V_{nl2}(I^-,H)$ from
Fig.~\ref{Figure2}a. There are two distinct features in these
curves. First, the signal is symmetric with respect to
simultaneous inversion of field and driving current, showing that
there is an easy vortex flow direction that is independent on
vortex polarity (vortices and antivortices flow easier in the same
direction). Second, additional periodic kinks become more
pronounced in the easy vortex flow direction.

The reason for the voltage asymmetry is due to the asymmetric
geometry of the structure - the vortex guiding channel is coupled
to the left and right endpoints through a different geometry. The
superconducting pads that couple to the vortex guiding channel
serve as vortex reservoirs that source vortices to the channel or
drain the vortices from the channel. On the left end we have the
funnel-shaped constriction and on the right end there is a square
superconducting pad. The ability for the vortices to enter into
the channel from these two pads is different; the funnel is
designed to jam the vortices as they enter the 1D channel. The
situation corresponds to the closed end channel described
in~\cite{vodolazov}. Jamming that occurs for vortices entering the
channel at the funnel end, causes reduction in vortex chain
mobility in the channel and the asymmetry of the $V_{nl}$ with
respect to the driving current. The same reduction will take place
for vortices of opposite polarity (antivortices) subject to
opposite driving current, as observed in Fig.~\ref{Figure3}. The
results are similar if one drives the vortices through the channel
by applying the Lorentz force with a supercurrent through the
left-most pair of contacts shown in the inset of
Fig.~\ref{Figure3}.

Furthermore, the mobility of the vortex chains in the channel will
be dependent on the matching of the vortex structure in the
source/drain pads and the channel. Matching of the vortex
configurations in the channel (vortex chains) to the vortex
configuration in the pads (vortex lattice) would cause higher
vortex mobility (larger $V_{nl}$ ) due to easier entrance/exit of
vortices, while non-matching configurations will cause reduction
in $V_{nl}$. In our experiments, when we are using the right pad
(square) as a source of vortices going into the channel, we
observe higher $V_{nl}$. We also observe more pronounced periodic
peaks in $V_{nl}$ corresponding to the geometrical matching effect
between the square vortex lattice in the pad and the vortex chain
structure in the channel. Further theoretical work could, possibly
correlate with these matching configurations.

In conclusion, we have studied the dynamics of non-local Abrikosov
vortex flow in mesoscopic superconducting Nb channels. High
rigidity of the vortex lattice in this material is responsible for
quasi-1D long-range correlated vortex movement. The magnetic field
dependence of the non-local voltage induced by the flux flow shows
that vortices form ordered vortex chains. Voltage asymmetry
(rectifying effect) with respect to the direction of vortex flow
is evidence that vortex jamming can significantly moderate vortex
dynamics in mesoscopic geometries. The finding can be applied to
future devices exploiting vortex dynamics and vortex
manipulation~\cite{semenov_prb2010}.

We would like to thank Ralu Divan (CNM, Argonne) for assistance in
preparation of the samples. This work as well as the use of the
Center for Nanoscale Materials and the Electron Microscopy Center
at Argonne National Laboratory were supported by UChicago Argonne,
LLC, Operator of Argonne National Laboratory (``Argonne'').
Argonne, a U.S. Department of Energy Office of Science laboratory,
is operated under Contract No. DE-AC02-06CH11357. M.I. would like
to acknowledge the support of U.S. Department of Energy under
Grant No. DE-SC0004556.

\bibliographystyle{apsrev}
\bibliography{Gbibliogrdet}

\end{document}